\documentclass[%
 reprint,
 amsmath,amssymb,
 aps,
]{revtex4-2}

\usepackage{graphicx}
\usepackage{dcolumn}
\usepackage{bm}
\usepackage[percent]{overpic}

\usepackage{xcolor}

\usepackage{amsthm}

\usepackage[colorlinks=true, linkcolor=blue, citecolor=blue, urlcolor=blue]{hyperref}

\newtheorem{theorem}{Theorem}

\begin{document}

\title{ Stability theory of flat band solitons in  nonlinear wave systems
}


\author{Cheng Shi}
\affiliation{ Department of Applied Physics and Applied Mathematics, Columbia University, New York, NY, 10027 USA}%

\author{Ross Parker}
\affiliation{IDA Center for Communications Research, Princeton, Princeton, NJ, 08540, USA}

\author{Panayotis G. Kevrekidis}
\affiliation{ Department of Mathematics and Statistics, University of Massachusetts Amherst, Amherst, 01003-4515, MA, USA}%

\affiliation{Department of Physics, University of Massachusetts Amherst, Amherst, 01003, MA, USA}

\affiliation{Theoretical Sciences Visiting Program, Okinawa Institute of Science and Technology Graduate University, Onna, 904-0495, Japan}

\author{Michael I. Weinstein}
\affiliation{ Department of Applied Physics and Applied Mathematics, Columbia University, New York, NY, 10027,  USA}%

\affiliation{ Department of  Mathematics, Columbia University, New York, NY, 10027 USA}

\date{\today}

\begin{abstract}
 We establish a sharp criterion for the stability of a class of compactly supported, homogeneous density
 ``minimal compact solitons'' or MCS states, of the time-dependent discrete nonlinear Schr\"odinger equation on a multi-lattice, 
 $\mathbb L$ ($\mathbb L$-DNLS). MCS states arise for multi-lattices where a nearest neighbor Laplace-type operator on $\mathbb L$ has a flat band. Our stability criterion is in terms of the explicit form of the nonlinearity and the projection of distinguished vectors onto the flat band eigenspace. We apply our general results to  MCS states of DNLS for the diamond, Kagom{\'e} and checkerboard lattices. 
In lattices where MCS states are unstable, 
we demonstrate how to engineer the nonlinearity to stabilize small amplitude MCS states. 
Finally, via systematic numerical computations, we put our analytical results in the context of global bifurcation diagrams.

\end{abstract}


\maketitle

\noindent{\bf Introduction:}
 There is significant interest in the dynamics of waves in naturally occurring and engineered crystalline media which exhibit a flat or nearly flat band in their band structure. A consequence of such spectra is the existence of localized, non-transporting and non-dispersing wavepackets or quasi-particles, a property which is used to enhance 
electron-electron  or light-matter  interactions in condensed matter \cite{BM2011,clugru:hal-03268886, masumoto2012exciton, whittaker2018exciton, kim2025real}, nonlinear effects in optics and photonics \cite{leykam2013flat,Leykam2018,vicencio2021photonic,rhim2021singular,danieli2024flat,guzman2014experimental,mukherjee2015observation,kajiwara2016observation, vicencio2015observation,weimann2016transport,xia2016demonstration, nguyen2018symmetry, ma2020direct} and metamaterial systems \cite{leykam2018artificial,wang2022observation,zhou2023observation,chasemayoral2024compact,wang2022observation,zhou2023observation,chasemayoral2024compact}. Among others, some of the experimental platforms in which these phenomena are investigated are:  laser-written optical waveguides, e.g.,  
in~\cite{guzman2014experimental,vicencio2015observation,mukherjee2015observation}, ultracold
atomic systems in periodic 
lattices~\cite{morsch,brazhnyi}
or coupled electrical circuits, e.g., in~\cite{chasemayoral2024compact,lape2025realizationcharacterizationallbandsflatelectrical}.

It is therefore particularly relevant to study the types of stable or long-lived localized excitations that can be supported in 
nonlinear wave models, 
whose underlying linear spectrum has a flat band.
A prototypical example arises in the discrete nonlinear Schr{\"o}dinger (DNLS) type wave systems arising in optics, superfluids and materials (see e.g. \cite{Lederer2008_PhysRep_DiscreteSolitons,Kevrekidis2009_DNLS,malomed2013spontaneous,AblowitzCole2022_PhysicaD_Review,granularBook}), on  the Kagom{\'e} lattice~\cite{Law}. Numerical studies reveal compactly supported nonlinear bound states from which one observes symmetry-breaking bifurcations ~\cite{vicencio2013discrete}. While there is substantial analytical theory of solitary wave stability, results on nonlinear dynamics of coherent structures in systems with linear flat band spectrum are mainly based on direct numerical simulations and linearized approximations \cite{Flach:2018}. A general, unifying theory of the dynamical stability of the nonlinear states of such systems, is of potential broad interest and our goal in this Letter is to present such a theory.

\begin{figure}[h]
   \centering
   \includegraphics[width=1.0\linewidth]{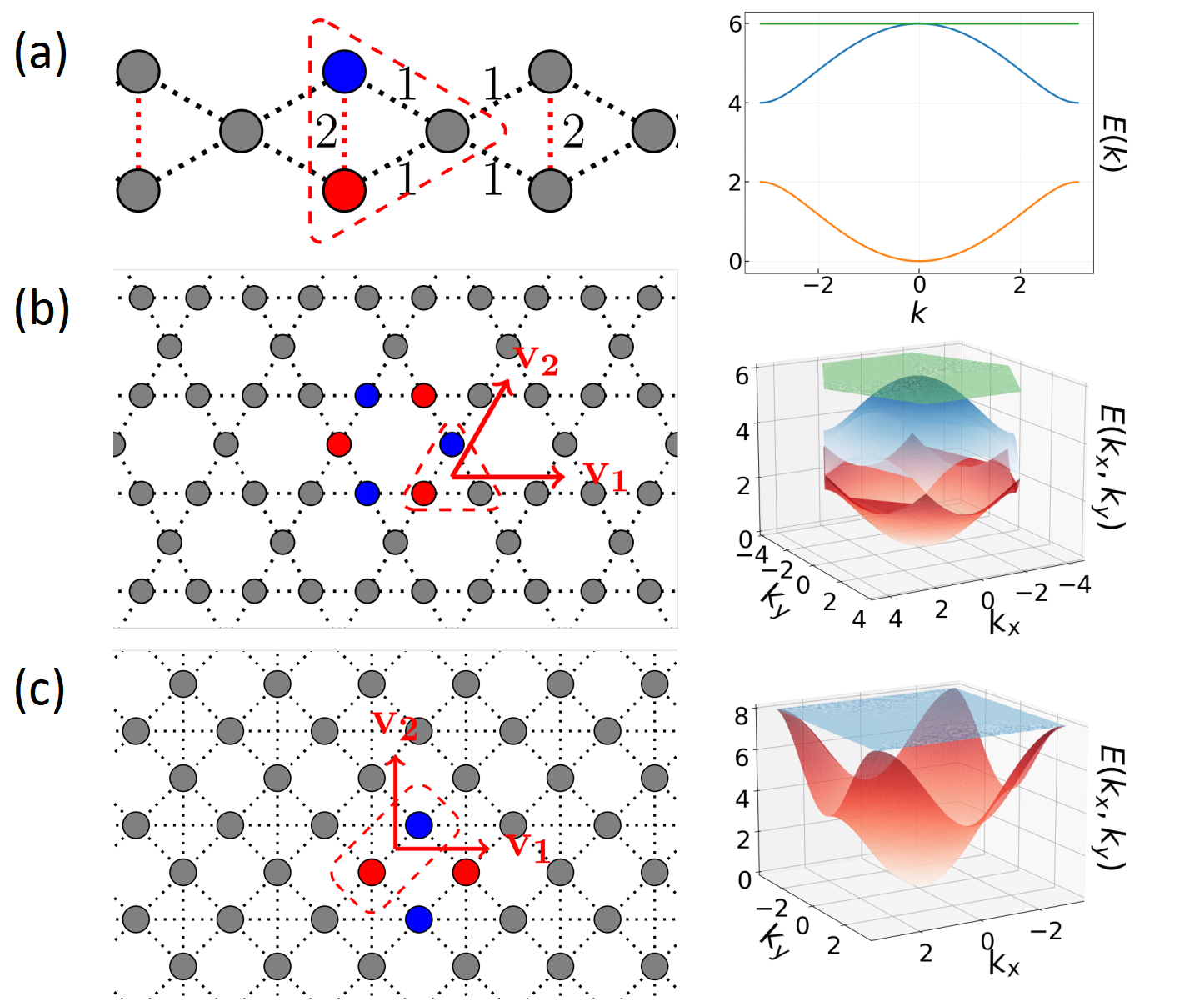}
    \caption{Lattices, $\mathbb L$, and band dispersion loci of $-\Delta_{\mathbb L}$ for three cases:
 (a) Diamond lattice ($\mathbb L = \mathbb D$) ; (b) Kagom{\'e} lattice ($\mathbb L = \mathbb K$); (c) Checkerboard lattice. ($\mathbb L = \mathbb Ch$). Vertices (``atoms'') comprising a fundamental cell are encircled by a dashed curve. Dotted lines connect nearest neighbors. Non-zero connectivity matrix elements, $w_{xy}$, are labeled for the lattice $\mathbb{D}$. For the lattices $\mathbb{K}$ and $\mathbb{C}h$, $w_{xy}\equiv1$. The flat band energy occurs at maximum energy of the spectrum of $-\Delta_{\mathbb L}$, $E=E_F$; for $\mathbb{D}$ and $\mathbb{K}$, $E_F = 6$, and for $\mathbb{C}h$, $E_F = 8$. Vertices in the support of a  minimal  compact flat band eigenstate of energy, $E_F$, are colored red and blue, which denote equal and opposite values of the wavefunction.}
    \label{fig:lattice presentation}
\end{figure}

More specifically, we develop a stability theory of symmetric ``minimal'' compactly supported homogeneous-density solitons, or simply ``minimal compact solitons'' (MCS states),  of the discrete nonlinear Schr\"odinger equation on a multi-lattice 
$\mathbb L$ ($\mathbb L-$DNLS).  Our lattice $\mathbb L$ is translation invariant and such that the nearest neighbor Laplacian type operator, $-\Delta_{\mathbb L}$, has a flat band at the maximum energy in its spectrum. Examples include the diamond, Kagom{\'e} and checkerboard lattices (Figure \ref{fig:lattice presentation}). 

{\it Our main result is a general sharp criterion for dynamical stability of minimal compact solitons.}
We explore the implications of this criterion in the context of the above three multi-lattice examples. For power-law nonlinearities, parametrized by a parameter $\sigma$, the bifurcation and stability properties of MCS states are described by two classes of bifurcation diagrams: (a) those for which MCS states are stable for powers, $\mathcal{N}<\mathcal{N}_\star(\sigma)$, a strictly positive threshold, followed by the bifurcation of a symmetry broken branch for $\mathcal{N}>\mathcal{N}_\star(\sigma)$ which destabilizes
the MCS, and (b) those for which the MCS states are unstable at any power. In the latter case, our general theory 
explains how to engineer the nonlinearity in order to stabilize MCS states at small amplitudes.  Via detailed numerical
bifurcation computations, we put our analytical results in the context of global bifurcation diagrams.
 
At the heart of our analysis is the observation that, for compactly supported states, the stability/instability analysis reduces to the spectral analysis of a linearized operator; in particular, 
the point 
spectrum or eigenvalues of a  finite rank perturbation of a discrete ``bulk" linear Hamiltonian with a flat band. Using the underlying discrete symmetries of the lattice and the subgroup of symmetries of the MCS states, we reduce this high-dimensional spectral problem to a family of spectral problems for rank one perturbations of the bulk operator acting in orthogonal symmetry subspaces. Strategies of this type have been employed to study symmetry induced band degeneracies; see  e.g. \cite{fefferman2012honeycomb,keller2018spectral,drouot_lyman:2025}. In particular, the 
relevant spectral problems are studied via asymptotic analysis of the resolvent (Green's function)  of the linearized operator for large spectral parameter and for the spectral parameter approaching the flat band energy. 

\noindent{\bf Laplacian on a multi-lattice, $\mathbb  L$ and Flat bands:} Fix a Bravais lattice, $\Lambda\subset\mathbb R^d_x$, its dual lattice $\Lambda^*\subset \mathbb R^d_k$ and Brillouin zone $\mathcal{B}\subset \mathbb R^d_k $. A multi-lattice, $\mathbb L$,  is a finite union of $N \geq 1$ 
of translated copies of $\Lambda$. The lattice $\mathbb L$ can be viewed as the union of $N-$atom cells. A {\it wave function} is a complex-valued function $\varphi$ defined on $\mathbb L$. 
Our Hilbert space is $l^2(\mathbb L)$, the space of complex-valued and square summable wave functions with its usual inner product and norm. We consider the Laplace-type periodic operators  $\Delta_{\mathbb{L}}$ on $\mathbb{L}$: 
\begin{equation}\label{eqn: Laplacian}
    \Delta_{\mathbb{L}}\varphi(x) = \sum_{y\sim x}w_{yx}(x)(\varphi(y) - \varphi(x)), \quad x\in\mathbb{L},
\end{equation}
where $\sum_{y\sim x}$ denotes a summation over all $y\in \mathbb L$ that are nearest neighbors of $x$. The connectivity matrix $(w_{xy})_{x,y\in\mathbb L}$ is taken to be real and symmetric, and $(x, y)\mapsto w_{xy}$ is $\Lambda\times\Lambda$-periodic. The operator $-\Delta_{\mathbb{L}}$ is bounded, non-negative and self-adjoint on $l^2(\mathbb L)$. Its band structure is given by the eigenpairs of $N-$ Floquet-Bloch eigenvalue maps: $E_1(k)\le E_2(k)\le\dots\le E_N(k)$, periodic with respect to $\Lambda^*$, with fundamental cell $\mathcal B$.  Figure \ref{fig:lattice presentation} displays $3$ multi-lattices with their band structures: the diamond lattice $\mathbb D$, the Kagom{\'e} lattice $\mathbb K$, and the checkerboard lattice $\mathbb Ch$.

The band structure contains a {\it flat band} if for some $j_0\in\{1,\dots,N\}$,  $E_{j_0}(k)=E_F$ is a constant for all $k\in\mathcal{B}$. In this paper, we focus on the setting where $-\Delta_{\mathbb L}$ has a single flat band at the upper limit, $E_F$, of its spectrum. The general properties of linear operators with flat bands have been studied extensively. For example, the spectral subspace of $-\Delta_{\mathbb L}$ with energy $E_F$ (``flat band eigenspace'') contains compactly supported states, and any flat band $l^2(\mathbb{L})$ eigenstate can be  approximated by such compactly supported states \cite{Kuchment2005}. 
Among these compactly supported states is a family of homogeneous density, alternating 
sign, compact states. 
This property can be used to obtain lattices with flat-bands via a line graph construction  \cite{Mielke1991, Mielke1991a, Kollar2020}. 

We are interested in {\it symmetric, minimal and homogeneous alternating compactly supported flat band states} or simply {\it minimal compact flat band states}. These are functions $\psi$ in the (maximal energy) flat band eigenspace of $-\Delta_{\mathbb L}$ such that the following two properties hold:
\begin{enumerate}
\item Symmetry: There is a symmetry $\mathfrak{S}\colon  \mathbb L \to \mathbb L$, cyclic of \textit{even} order $s$, such that $w_{\mathfrak{S}x,\mathfrak{S}y}=w_{xy}$,  and a point $x_0\in\mathbb L$ such that 
${\rm supp} (\psi)$ consists of the $s$ distinct points: $x_0, \mathfrak{S}x_0,\dots,\mathfrak{S}^{s-1}x_0$. ${\rm Orb_{\mathfrak{S}}}(x_0)$ is called  the orbit of $x_0$,   under  $\mathfrak S$.
\item Alternating: $\psi$ has uniform amplitude along its support, and $\psi(x) = a\sum_{j=0}^{s-1} (-1)^j \delta_{{\mathfrak S}^j x_0}(x)$ for some $a\neq 0$, and $\delta_{y}(x)$ denotes the Kronecker delta function at $y$.   
\item Minimality: If $-\Delta_{\mathbb L}e=E_Fe$ and ${\rm supp}(e)\subset {\rm supp}(\psi)$, then $e\in{\rm span}\{\psi\}$.
\end{enumerate}

The colored vertices in each lattice of Figure \ref{fig:lattice presentation} mark the support of minimal compact flat band states for the diamond lattice ($d=1$, $N=3$; $\mathfrak{S}_{\mathbb{D}}=$  horizontal reflection, $s=2$; $\psi=$ dipole), the Kagom{\'e} ($d=2$, $N=3$; $\mathfrak{S}_{\mathbb{K}}$ is $\pi/6$-rotation, $s=6$; $\psi=$ hexagonal) and the checkerboard lattice ($d=2$, $N=2$; $\mathfrak{S}_{\mathbb{C}h}$ is $\pi/4$-rotation, $s=4$;  $\psi=$ square). 
Any $\Lambda-$translate of these states is also a minimal compact flat band state and the set of all its lattice translates is a dense subset of the $l^2-$ nullspace of $-\Delta_{\mathbb L}-E_F$ \cite{SPKW-inpreparation}. 

\noindent{\bf Discrete nonlinear Schr\"odinger equation:}
We consider the discrete nonlinear Schrodinger equation (DNLS)~\cite{Kevrekidis2009_DNLS} on the lattice $\mathbb{L}$ governing  $\Psi\in l^2(\mathbb{L};\mathbb{C})$:
\begin{equation}\label{eqn: DNLS}
    i\partial_t\Psi(x, t) = -\Delta_{\mathbb{L}}\Psi(x, t) + f(|\Psi(x,t)|^2)\Psi(x, t), 
\end{equation}
 $\mathbb L-$DNLS is a  Hamiltonian system, $i\partial_t\Psi=\delta \mathcal{H}/\delta \overline{\Psi}$. The time-invariant Hamiltonian is given by:
\begin{equation}\label{eqn: Hamiltonian energy}
    \mathcal{H}[\Psi,\overline\Psi] := \langle -\Delta_{\mathbb{L}}\Psi(\cdot, t), \Psi(\cdot, t)\rangle_{l^2(\mathbb{L})} + \sum_{x\in\mathbb{L}}F(|\Psi(x, t)|^2),
\end{equation}
where $F(y) = \int_0^y f(z)\,dz$, and the squared $l^2$-norm, also known as the {\it power} of the state $\Psi$ is:
\begin{equation}\label{eqn: power}
    \mathcal{N}[\Psi] :=  \sum_{x\in \mathbb{L}}|\Psi(x, t)|^2. 
\end{equation}
A standing wave solution is of the form $e^{-iE_{\rm nl}t}\psi(x)$, where $-\Delta_{\mathbb L}\psi + f(|\psi|^2)\psi = E_{\rm nl}\psi$ and $\psi\in l^2(\mathbb L)$. 

Since the nonlinear term in 
\eqref{eqn: DNLS} is onsite, any minimal compact flat band state, arising from the flat band of $-\Delta_{\mathbb{L}}$, extends to a nonlinear bound state, a ``minimal compact soliton'' or MCS state, of $\mathbb{L}$-DNLS \eqref{eqn: DNLS}: 
\begin{equation}\label{eqn: nonlinear frequency}
 \psi_a e^{-iE_{\rm nl}t} = a\ \sum_{j=1}^s (-1)^j\delta_{\mathfrak S^j x_0}(x)\ e^{-iE_{\rm nl} t}, 
\end{equation}
where $E_{\rm nl}=E_{\rm nl}(a) = E_F + f(a^2)$ is an amplitude-dependent nonlinear frequency, and by \eqref{eqn: nonlinear frequency} we have
\begin{equation}
    -\Delta_{\mathbb L}\psi_a + f(a^2)\psi_a = E_{\rm nl}\psi_a,\quad \psi\in l^2(\mathbb L).
\label{eq:psi-eqn-a}\end{equation}

In this article we consider {\it defocusing} or {\it repulsive}  nonlinearities of power law type: $f(|x|^2) = |x|^{2\sigma}$, $F(|x|^2)= |x|^{2\sigma+2}/(\sigma+1)$~\cite{CUEVAS200967,Kevrekidis2009_DNLS}.
For defocusing nonlinearities, the bifurcation of compactly supported nonlinear bound states is toward frequencies above the linear flat band spectrum, $E_{\rm nl}(a)>E_F$.
Our main result, Theorem \ref{theorem: power law nonlinearity stability}, characterizes when the minimal compact solitons are dynamically stable and when they are unstable. {A universal character of the bifurcation diagram emerges.

\begin{theorem}\label{theorem: power law nonlinearity stability}
Assume $-\Delta_{\mathbb L}$ has a flat band at the maximum of its spectrum. Consider $\mathbb{L}$-DNLS with power-law defocusing nonlinearity $f(|x|^2) = |x|^{2\sigma}$. 
Let $\nu\mapsto \psi^\nu$ denote a family of minimal compact solitons, parametrized by $\mathcal{N}[\psi^\nu]=\nu$.
 
  There exists a critical value of the nonlinearity parameter $0 < \sigma^{\rm cr}_{\mathbb L}\leq \infty$, such that: 
    \begin{enumerate}
        \item [(1)]
        for every $\sigma\in(0, \sigma^{\rm cr}_{\mathbb L})$, there is a threshold power $\mathcal{N}_{\rm thr}(\sigma)$, with  $0 < \mathcal{N}_{\rm thr}(\sigma) < \infty$, such that:\\
        (i) $\psi^\nu$ is dynamically stable if $\nu < \mathcal{N}_{\rm thr}(\sigma)$, and \\ (ii) $\psi^\nu$ is unstable if $\nu > \mathcal{N}_{\rm thr}(\sigma)$. 
        \item [(2)]
        for every $\sigma > \sigma^{\rm cr}_{\mathbb L}$, $\psi^\nu$ is  dynamically unstable for every $\nu>0$. 
    \end{enumerate}
    The stability / instability transitions corresponding to (i) and (ii) are represented in the schematic of Figure \ref{fig:local bifurcation}.
\end{theorem}
 \noindent An extension of the current theory to general nonlinearities, $f$, will be presented in a forthcoming article \cite{SPKW-inpreparation}.

\begin{figure}[h]
    \centering
    \begin{overpic}[width=0.49\linewidth]{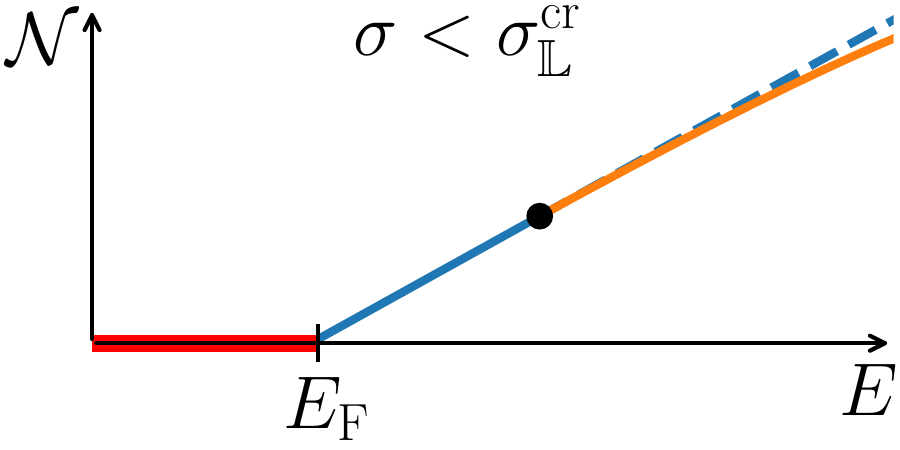}
        \put(3,53){\textbf{(a)}} 
    \end{overpic}%
    \hfill
    \begin{overpic}[width=0.49\linewidth]{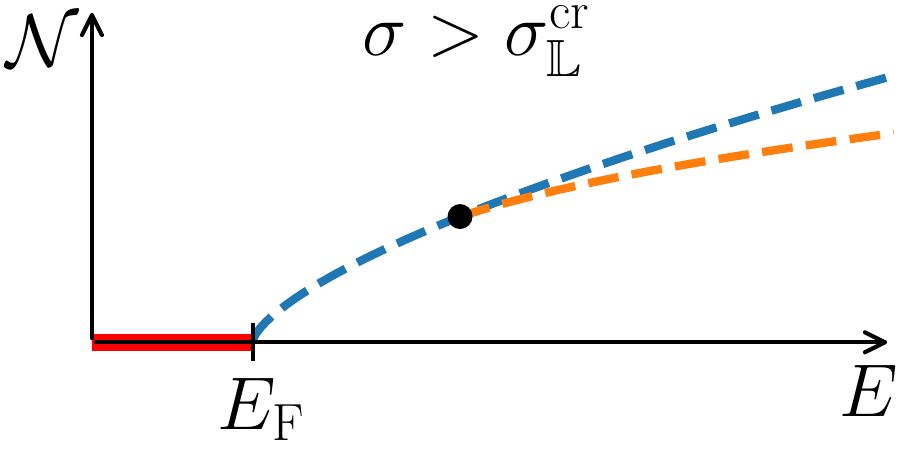}
        \put(3,53){\textbf{(b)}}
    \end{overpic}
    \caption{
    Universal character of stability transitions for minimal compact solitons (MCS states),  which bifurcate from a flat band at $E=E_F$. Theorem \ref{theorem: power law nonlinearity stability} implies two possible scenarios: (a) when $\sigma < \sigma^{\rm cr}_{\mathbb L}$, the compact state is at first stable and then transfers its stability to a second branch in a symmetry breaking bifurcation, and (b) when $\sigma > \sigma^{\rm cr}_{\mathbb L}$, the compact state is always unstable.  
    In general, the bifurcation can be sub- or super-critical. 
    }
    \label{fig:local bifurcation}
\end{figure}
%
%


 We next discuss the notions of stability and instability in Theorem \ref{theorem: power law nonlinearity stability},  and the ideas underlying the proof.


\noindent{\bf Stability and instability:} Our treatment of stability and instability draws inspiration from studies of the stability of ground state solitary waves of the focusing nonlinear Schr\"odinger equation, e.g.  \cite{Weinstein1986,Grillakis1987,grillakis1988,Weinstein1999,Weinstein2015}. Consider a homogeneous and compactly supported nonlinear bound state, $\psi^\nu$. The set of all phase translations of $\psi^\nu$ is the {\it orbit of $\psi^\nu$}:  $\mathcal{O}_{\psi^\nu} := \left\{\psi^\nu e^{i\theta}\colon \theta\in [0, 2\pi)\right\}$. The appropriate notion of stability is {\it nonlinear orbital stability};  $\psi$ is orbitally stable if ${\rm dist}(\Psi(\cdot, t=0), \mathcal{O}_\psi)$ small implies that ${\rm dist}(\Psi(\cdot, t), \mathcal{O}_\psi)$ is small for every $t \ne0$. Here,  ${\rm dist}\big(\Psi(\cdot,t), \mathcal{O}_{\psi^\nu}\big)$ is the minimal $l^2$-distance from the point $\psi(\cdot,t)$ to the set $\mathcal{O}_{\psi^\nu}$.

The general strategy for determining the orbital stability or instability of a nonlinear bound state is outlined in the End Matter. The analysis reduces to the spectral properties of two linear
 operators, $L_+$ and $L_-$, which arise in the linearized DNLS evolution governing small perturbations. In the case of MCS states, orbital stability is equivalent to the existence of exactly one non-negative eigenvalue in $L_+$. 

\noindent {\bf Stability/instablity of minimal compact solitons:}
In the case of power nonlinearities, if $\psi_a$ is an MCS state of amplitude $a$ (see \eqref{eqn: nonlinear frequency}),  then  $f(|\psi_a|^2)=|a|^{2\sigma}\delta_{{\rm supp}(\psi_a)}$, where the subscript denotes
the support of the wavefunction $\psi_a$. The operators $L_\pm$ are then:
\begin{subequations}
\label{eq:Lpm}
 \begin{align}
  L_+ &= -\Delta_{\mathbb{L}} - E_{\rm nl}(a)  + (2\sigma + 1)|a|^{2\sigma}\delta_{{\rm supp}(\psi)}\\
  L_- &= -\Delta_{\mathbb{L}} - E_{\rm nl}(a) + |a|^{2\sigma}\delta_{{\rm supp}(\psi)} \ ,   
 \end{align}
 \end{subequations}
 where $E_{\rm nl}(a)$  is displayed just after \eqref{eqn: nonlinear frequency}.
 We next discuss the spectra of $L_-$ and $L_+$.
 
First, $L_-$, as an operator on $l^2(\mathbb{L})$, is negative semi-definite with $0$ as its maximum eigenvalue. Indeed,  both $-\Delta_{\mathbb{L}} - E_F$ and $|a|^{2\sigma}\delta_{{\rm supp}(\psi)} - |a|^{2\sigma}$ are negative semi-definite, and, by \eqref{eq:psi-eqn-a}, $L_-\psi_a=0$.  On the other hand, $L_+$,  as an operator on $l^2(\mathbb{L})$, is indefinite. In fact, 
 by \eqref{eq:psi-eqn-a}, $L_+$ has a strictly positive eigenvalue: $L_+\psi_a=2\sigma |a|^{2\sigma}\psi_a$. 

In view of the above spectral criterion on $L_+$, our strategy for assessing the stability or instability of $\psi_a$ is to, for a fixed nonlinearity parameter $\sigma$,  study the number of positive eigenvalues of $L_+$, $n_+(a;\sigma)$, as the amplitude parameter $a$ varies; when $n_+(a;\sigma)>1$, the state $\psi_a$ is unstable. Further, if $n_+(a;\sigma)=1$, then the above variational considerations imply nonlinear orbital stability.
 
\noindent{\bf Analysis of $L_+$ as $a$ varies:} Since $\psi_a$ is supported on exactly $s$ sites of $\mathbb L$, the operator $L_+$ is a  perturbation of the bulk operator $-\Delta_\mathbb{L} - E_{\rm nl}$ of finite rank equal to $s$. By the min-max principle of self-adjoint operators, the spectrum of $L_+$
has at most $s$ eigenvalues above $E_F-E_{\rm nl}(a)=-|a|^{2\sigma}<0$,  the upper limit of its continuous spectrum. As noted above,  $L_+$ has a strictly positive eigenvalue and hence, for the state $\psi_a$ to be stable, it is necessary that the remaining $\le s-1$ eigenvalues of $L_+$ are strictly smaller than zero. 

Studying the eigenvalues of $L_+$ which cross zero energy as $a$ varies, via perturbation theory about the bulk operator:
\begin{equation}\label{eq:Lbulk-a}
  L_{{\rm bulk},a}= -\Delta_{\mathbb L}-E_{\rm nl}(a)=-\Delta_{\mathbb L}-E_F-|a|^{2\sigma}. 
\end{equation}
is subtle since, as $a\to0^+$,  $E_{\rm nl}(a)\to E_F$,  an eigenvalue of infinite multiplicity of $-\Delta_{\mathbb L}$. 

 \noindent{\bf Linear algebra and symmetry:} Point  spectra (eigenvalues) of $L_+$ are poles of the resolvent operator $\lambda\mapsto (L_+-\lambda I)^{-1}$ on the real axis and outside the continuous spectrum of $L_+$.  At the core of our analysis lies
 the fact that the determination of the point spectrum can be reduced, via the Sherman-Morrison-Woodbury identity \cite{Golub-VanLoan2013}, to an analysis of the zeros of the determinant of an  $s$-by-$s$ matrix, depending analytically on $\lambda$ and on the amplitude $a$. 
 A detailed analysis is possible by taking advantage of the symmetries of the lattice $\mathbb L$ and of our MCS states. 
 
Let $U_{\mathfrak{S}}$ be the symmetry on $l^2(\mathbb L)$ induced by $\mathfrak{S}$: $U_{\mathfrak S}f(x) = f(\mathfrak{S}^{-1}x)$. Since $U_\mathfrak{S}$ is a symmetry of order equal to $s$, we have $U_\mathfrak{S}^s={\rm Id}$, and thus its eigenvalues are the $s$-roots of unity $\{\omega^n\}_{n=0}^{s-1}$, where $\omega = e^{2\pi i/s}$. Further, $U_\mathfrak{S}$ is a normal operator, so we may decompose our Hilbert space $l^2(\mathbb L)$ into an orthogonal sum of $s$ invariant eigenspaces, $l_{\omega^n}^2(\mathbb{L})$, of $U_\mathfrak{S}$: $l^2(\mathbb{L})=\bigoplus_{n = 0}^{s-1}l_{\omega^n}^2(\mathbb{L})$. 
Second, observe that  $U_\mathfrak{S}$ commutes with $-\Delta_{\mathbb L}$ and 
multiplication by $|a|^{2\sigma}\delta_{{\rm supp}(\psi_a)}$, and so $U_\mathfrak{S}$ commutes with $L_+$. 
The action of $L_+$ on $l^2(\mathbb L)$ can therefore be decomposed into those of $s$-independent operators $L_{+,\omega^n}$, each defined as $L_+$ restricted to $l_{\omega^n}^2(\mathbb{L})$. Note that the minimality property of the MCS state $\psi_a$ implies that $s$, the order of the symmetry, is equal to the rank of the perturbation $L_+-L_{\rm bulk,a}$.

Next, we represent  
$\delta_{{\rm supp}(\psi_a)} = \sum_{n=0}^{s-1}e_n \langle e_n,\cdot\rangle$, where  $e_n\in l^2_{\rm \omega^n}(\mathbb L)$, ${\rm supp}(e_n)={\rm supp}(\psi_a)$, for $n=0,1,\dots,s-1$, and the set $\{e_n\}_{n=0}^{s-1}$ to be an orthonormal subset of $l^2(\mathbb L)$. Thus,   $\delta_{{\rm supp}(\psi)}|_{l^2_{\omega^n}(\mathbb{L})} = e_n\langle e_n,\cdot\rangle$ is a rank-one operator $l_{\omega^n}^2(\mathbb{L})$; see the End Matter for further discussion. Therefore, for each $n=0,\dots,s-1$,  
\begin{equation}\label{eq:L+rank1} L_{+,\omega^n}= -\Delta_{\mathbb L}-E_{\rm nl}(a) +(2\sigma+1)|a|^{2\sigma}e_n \langle e_n,\cdot\rangle,\end{equation} a rank one perturbation of $L_{{\rm bulk},a}$ on $l_{\omega^n}^2(\mathbb L)$, given in \eqref{eq:Lbulk-a}. 
Moreover, since $-\Delta_{\mathbb L}-E_F$ is negative semi-definite, each $L_{+,\omega^n}$ has at most one eigenvalue, denoted $\lambda_n(|a|^{2\sigma})$, above energy $-|a|^{2\sigma}$, the upper edge of the continuous  spectrum of $L_{+,\omega^n}$.  
%
Note that, reflecting its oscillatory nature, 
$\psi_{a}$ itself is a multiple of $e_{s/2}$, and is an eigenstate of $L_{+, -1}$, with eigenvalue $2\sigma |a|^{2\sigma}$. 

At this point, we have reduced the problem of determining the point spectrum of $L_+$ to tracking the $s-1$ simple eigenvalues of the family of operators $\{L_{+,\omega^n}\}_{n\neq s/2}$ as the MCS amplitude varies. 
Using the Sherman-Morrison-Woodbury identity \cite{Golub-VanLoan2013} we can represent the resolvent of $L_{+,\omega^n}$ as a perturbation of the resolvent of $L_{\rm{bulk, a}}$; see the End Matter.  For each $n = 0,\cdots, s-1$, the eigenvalue of $L_{+,\omega^n}$, being pole of its resolvent, is given by the zero, $\lambda_n(|a|^{2\sigma})$, of the scalar function: 
\begin{equation}\label{equation: Sherman-Morrison --- invertibility}
F_n( \lambda,|a|^{2\sigma}) := 1 + (2\sigma + 1)|a|^{2\sigma}\langle e_n, R_{{\rm bulk},a}(\lambda) e_n\rangle, 
\end{equation}
where $R_{{\rm bulk},a}(\lambda) =(L_{{\rm bulk},a}-\lambda )^{-1}$. 
We study $\lambda_n(|a|^{2\sigma})$ as $a$ varies. An asymptotic analysis of $F_n(\lambda, |a|^{2\sigma})$, for $|a|^{2\sigma}\to \infty$ and $|a|^{2\sigma}\downarrow0$,  detailed in the End Matter, yields the sharp stability criterion: 
\begin{equation}\label{equation: stability criterion} 
\max_{n\neq s/2}\|P_{E_F}^{-\Delta_{\mathbb{L}}}e_n\|_2^2 < \frac{1}{2\sigma + 1},
\end{equation}
where $P_{E_F}^{-\Delta_{\mathbb{L}}}$ denotes the $-\Delta_{\mathbb{L}}$-spectral projection operator onto the flat band at energy $E_F$.

If condition (\ref{equation: stability criterion}) is satisfied, then there exists $a_\star>0$ such that for $0 < a < a_\star$, the $s-1$ eigenvalues $\lambda_n(|a|^{2\sigma})$, with $n\neq s/2$, are strictly negative, and the MCS state of this amplitude is stable. On the other hand, if $a > a_\star$, then at least one of $\lambda_n(|a|^{2\sigma})$ with $n\neq s/2$ becomes positive and we have instability due to the existence of more than one positive eigenvalue of $L_+$. 
A typical plot of the motion of $\lambda_n(|a|^{2\sigma})$ under the scenario of condition \eqref{equation: stability criterion} is given in Figure \ref{fig:Sherman-Morrison}. 

If the condition (\ref{equation: stability criterion}) is violated, then for every $a$, there is at least one $n\neq s/2$ such that $\lambda_n(|a|^{2\sigma}) > 0$, implying a second positive eigenvalue of $L_+$ along with $2\sigma |a|^{2\sigma}$. In conclusion:
 {\it there is a range of amplitudes, $0<a<a_\star(\sigma)$, of MCS  (with corresponding power range, $0<\mathcal{N}<\mathcal{N}_\text{thr}(\sigma)$) which are orbitally stable if and only if the spectral condition \eqref{equation: stability criterion} holds.}
 
Suppose condition \eqref{equation: stability criterion} is satisfied. As the amplitude increases, each $\lambda_n(|a|^{2\sigma})$ ($n\ne s/2$) will eventually cross zero. The MCS state $\psi_a$ loses its stability at the first such zero-crossing; see panel (a) of Figure \ref{fig:local bifurcation}. Coincident with each successive crossing, is a new symmetry-broken state which  bifurcates from the branch $(\psi_a, E_{\rm nl}(a))$; see the End Matter discussion of global bifurcations. 

To conclude the discussion of Theorem 1, we argue that for generic lattices there is always a $\sigma_{\mathbb L}^{\rm cr}>0$.  By the minimality of MCS states,  $0 \leq \|P_{E_F}^{-\Delta_{\mathbb{L}}}e_n\|_2^2 < 1$, i.e., $e_n$ is not a flat-band eigenstate, for every $n\neq s/2$. This condition is independent of the nonlinearity parameter $\sigma$.  
Hence, for small enough $\sigma > 0$, \eqref{equation: stability criterion} is always guaranteed to be satisfied,
therefore establishing the existence of $\sigma_{\mathbb{L}}^{\rm cr} > 0$.

\begin{figure}[h]
    \centering
    \includegraphics[width=1.0\linewidth]{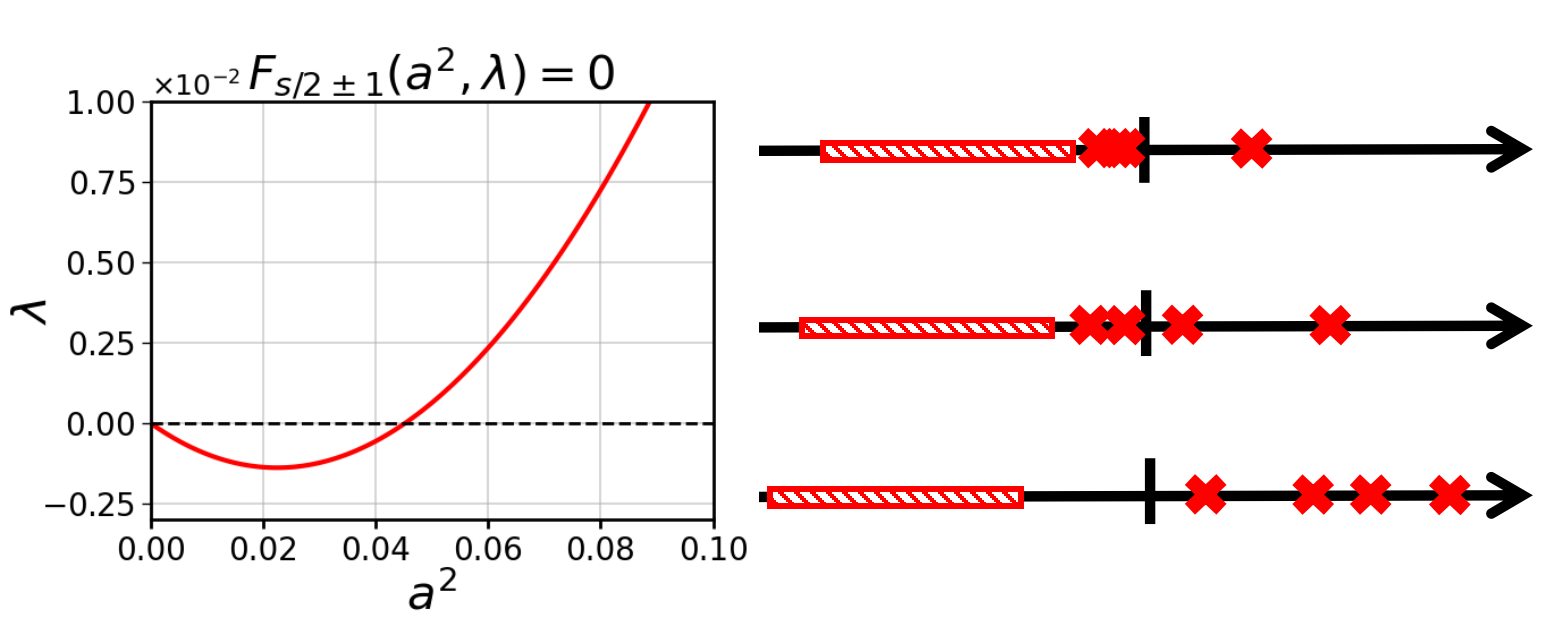}
    \caption{{Motion of eigenvalues, $\lambda_n(|a|^2)$, Kagom{\'e} case. Right: Schematic showing $|a|$ small ($\mathcal{N}[\psi_a]$ small), $L_+=\bigoplus_{n\ne s/2}L_{+,\omega^n}$ has $s-1$ strictly negative eigenvalues in the interval $(-|a|^{2\sigma},0)$. As $a$ increases, the discrete eigenvalue of  each $L_{+,\omega^n}$, zero of $\lambda\mapsto F_n(\lambda,|a|^{2\sigma})$, crosses zero energy from negative to positive values, and (may) induce successive bifurcations. The minimal compact soliton $\psi_a$ becomes unstable at the first such crossing. Left: Trajectory of $\lambda_{\frac{s}{2}\pm 1}(|a|^2)$ of $L_+$ as $a>0$ varies for cubic nonlinearity $\sigma = 1$.}}
    \label{fig:Sherman-Morrison}
\end{figure}

\noindent{\bf Application $\mathbb D$, $\mathbb K$ and $\mathbb Ch$ lattices:} 
We now discuss the implications of Theorem \ref{theorem: power law nonlinearity stability} for the $\mathbb D$, $\mathbb K$ and $\mathbb Ch$ lattices. Recall that our defocusing nonlinearity is $|\psi|^{2\sigma}\psi$. Theorem 1 ensures that if $\sigma<\sigma_{\mathbb L}^{\rm cr}$ then all sufficiently small amplitude MCS states are stable (Figure \ref{fig:local bifurcation}, left panel) and if  $\sigma>\sigma_{\mathbb L}^{\rm cr}$ all MCS states are unstable (Figure \ref{fig:local bifurcation}, right panel).  For $\mathbb D$, $\mathbb K$ and $\mathbb Ch$ lattices, 
our explicit calculation
based on~(\ref{equation: Sherman-Morrison --- invertibility}) yields that 
the critical nonlinearity values  are:
 $\sigma^{\rm cr}_{\mathbb D} = \infty$, $\sigma^{\rm cr}_{\mathbb K}\approx 1.268$ and $\sigma^{\rm cr}_{\mathbb Ch}\approx 0.829$. Theorem \ref{theorem: power law nonlinearity stability} implies for the cubic nonlinearity ($\sigma=1$) that $\mathbb D-$ and $\mathbb K-$DNLS states are stable below a strictly positive
$l^2-$ threshold 
($\nu^\star_{\mathbb D}\approx 2.61$ and  $\nu^\star_{\mathbb K}\approx 0.19$). In fact, for the diamond lattice there is a transition between stable states, at sufficiently small amplitude, and unstable states above some threshold amplitude for any nonlinearity parameter $\sigma$ (since $\sigma^{\rm cr}_{\mathbb D} = \infty$). 
For $\mathbb K-$DNLS, when $\sigma>\sigma_{\mathbb K}^{\rm cr}\approx 1.268$, all MCS states are unstable. 
Consider now $\mathbb Ch-$DNLS, with $\sigma=1$. Here,  MCS states of arbitrary $l^2$ norm are unstable. However, for $\sigma<\sigma^{\rm cr}_{\mathbb Ch}\approx 0.829$ there is a transition between stable sufficiently small amplitude states and unstable states.

Our analysis provides information about the nature of symmetry breaking bifurcations from the branch of MCS states, regardless of their stability.  As the amplitude, $a$, is increased,  successive simple eigenvalues of $L_{+,\omega^n}$, for some $n\ne s/2$ pass through zero energy, inducing a bifurcation. The zero energy eigenspace gives an infinitesimal view of the manner in which each bifurcating branch breaks the symmetries of the  MCS branch. Figure~\ref{fig:BD}, based on detailed numerical computations, supports Theorem 1, and also 
shows a global zoomed out view of the various localized states of $\mathbb L$-DNLS for $\mathbb L=\mathbb D, \mathbb K$ and $\mathbb Ch$. For further discussion, see the End Matter.


\noindent{\bf Summary and discussion:} 
 We have established a sharp criterion for the stability of MCS states of the time-dependent discrete defocusing nonlinear Schr\"odinger equation on a multi-lattice $\mathbb L$ ($\mathbb L$-DNLS), and have demonstrated our results for the 
 diamond, Kagom{\'e} and checkerboard lattices. For the class of power nonlinearities, we have shown that by adjusting the nonlinearity parameter to be sufficiently small, we can always stabilize the small amplitude MCS states. 

 We expect the variational, spectral theoretic and symmetry ideas of this article to play a central role in further studies of these
 and related lattices. For example, (1) flat bands may appear at energies which are interior to the band spectrum. This occurs, for example, in the Lieb lattice, for which DNLS has MCS states; see also the experimental study of a class of diamond lattices \cite{chasemayoral2024compact}.  Are such  MCS states unstable due to coupling to radiation modes? 
(2) Here we considered the defocusing
nonlinearity, but a natural next step is
to complement such studies with a 
detailed analysis of the focusing case.
 (3) Another extension of the current study 
 is to lattices arising as  line graphs of an underlying lattice \cite{Kollar2020}. The  present paper covers the 
 Kagom{\'e} lattice, which is the line graph of the honeycomb lattice
 and the Checkerboard lattice, which is the line graph of $\mathbb Z^2$. Other examples are the line graph of the triangular lattice or the pyrochlore~\cite{Huang2024_pyrochlore_flatbands}, which is the line graph of the 3D diamond lattice. Extensions
 to such 3D settings would be of 
 particular interest.

{\it Acknowledgements.}
This research was supported by the U.S. National Science Foundation under the awards DMS-2204702 and PHY-2408988 (PGK). This research was partly conducted while P.G.K. was 
visiting the Okinawa Institute of Science and
Technology (OIST) through the Theoretical Sciences Visiting Program (TSVP). 
This work was also 
supported by a grant from the Simons Foundation
[SFI-MPS-SFM-00011048, P.G.K]. 
P.G.K. also acknowledges discussions with Prof. R. Carretero-Gonz{\'a}lez at
an early stage of this work.
M.I.W. and C.S. were supported in part by NSF grants: DMS-1908657, DMS-1937254 and DMS-2510769, and Simons Foundation Math+X Investigator Award \#376319 [M.I.W.]. Part of this research was carried out during the 2023-24 academic year, when M.I.W. was a Visiting Member in the School of Mathematics, Institute of Advanced Study, Princeton, supported by the Charles Simonyi Endowment, and a Visiting Fellow in the Department of Mathematics at Princeton University.

\bibliographystyle{unsrt}  
\bibliography{references}  

@PREAMBLE{
 "\providecommand{\noopsort}[1]{}" 
 # "\providecommand{\singleletter}[1]{#1}%" 
}

@book{granularBook,
  author    = {C.~Chong and P. G.~Kevrekidis},
  title     = {Coherent Structures in Granular Crystals: From Experiment and Modelling to Computation and Mathematical Analysis},
  publisher = {Springer},
  address   = {New York},
  year      = {2018}
}

@article{Huang2024_pyrochlore_flatbands,
  author    = {Huang, Jianwei and Setty, Chandan and Deng, Liangzi and You, Jing-Yang
               and Liu, Hongxiong and Shao, Sen and Oh, Ji Seop and Guo, Yucheng
               and Zhang, Yichen and Yue, Ziqin and Yin, Jia-Xin and Hashimoto, Makoto
               and Lu, Donghui and Gorovikov, Sergey and Dai, Pengcheng
               and Denlinger, Jonathan D. and Allen, J. W. and Hasan, M. Zahid
               and Feng, Yuan-Ping and Birgeneau, Robert J. and Shi, Youguo
               and Chu, Ching-Wu and Chang, Guoqing and Si, Qimiao and Yi, Ming},
  title     = {Observation of flat bands and Dirac cones in a pyrochlore lattice superconductor},
  journal   = {npj Quantum Materials},
  year      = {2024},
  volume    = {9},
  articleno = {71},
  doi       = {10.1038/s41535-024-00683-x},
  url       = {https://www.nature.com/articles/s41535-024-00683-x}
}

@article{CUEVAS200967,
title = {Discrete solitons in nonlinear Schrödinger lattices with a power-law nonlinearity},
journal = {Physica D: Nonlinear Phenomena},
volume = {238},
number = {1},
pages = {67-76},
year = {2009},
issn = {0167-2789},
doi = {https://doi.org/10.1016/j.physd.2008.08.013},
url = {https://www.sciencedirect.com/science/article/pii/S0167278908003102},
author = {J. Cuevas and P.G. Kevrekidis and D.J. Frantzeskakis and B.A. Malomed}
}

@article{Flach:2018,
title = {Compact discrete breathers on flat-band networks},
journal = {Low Temperature Physics},
volume = {44},
number = {7},
pages = {678-687},
year = {2018},
author = {C. Danieli and A. Maluckov and S. Flach}
}

@article{Lederer2008_PhysRep_DiscreteSolitons,
  author  = {Lederer, Falk and Stegeman, George I. and Christodoulides, Demetri N. and Assanto, Gaetano and Segev, Mordechai and Silberberg, Yaron},
  title   = {Discrete solitons in optics},
  journal = {Physics Reports},
  year    = {2008},
  volume  = {463},
  number  = {1--3},
  pages   = {1--126},
  doi     = {10.1016/j.physrep.2008.04.004},
  url     = {https://www.sciencedirect.com/science/article/pii/S0370157308001257}
}

@book{Kevrekidis2009_DNLS,
  author    = {Kevrekidis, Panayotis G.},
  title     = {The Discrete Nonlinear Schr{\"o}dinger Equation: Mathematical Analysis, Numerical Computations and Physical Perspectives},
  series    = {Springer Tracts in Modern Physics},
  volume    = {232},
  publisher = {Springer},
  address   = {Berlin, Heidelberg},
  year      = {2009},
  doi       = {10.1007/978-3-540-89199-4},
  isbn      = {978-3-540-89198-7},
  url       = {https://link.springer.com/book/10.1007/978-3-540-89199-4}
}

@article{AblowitzCole2022_PhysicaD_Review,
  author    = {Ablowitz, Mark J. and Cole, Justin T.},
  title     = {Nonlinear optical waveguide lattices: Asymptotic analysis, solitons, and topological insulators},
  journal   = {Physica D: Nonlinear Phenomena},
  year      = {2022},
  volume    = {440},
  pages     = {133440},
  doi       = {10.1016/j.physd.2022.133440},
  url       = {https://doi.org/10.1016/j.physd.2022.133440},
  note      = {Review article},
  eprint    = {2212.11993},
  archivePrefix = {arXiv},
  primaryClass  = {nlin.PS}
}

@article{brazhnyi,
author = {Brazhnyi, V. A. and Konotop, V. V.},
title = {Theory of nonlinear matter waves
        in optical lattices},
journal = {Modern Physics Letters B},
volume = {18},
number = {14},
pages = {627-651},
year = {2004},
doi = {10.1142/S0217984904007190},

URL = { 
    
        https://doi.org/10.1142/S0217984904007190
    
    

},
eprint = { 
    
        https://doi.org/10.1142/S0217984904007190
    
    

}
}

@article{morsch,
  title = {Dynamics of Bose-Einstein condensates in optical lattices},
  author = {Morsch, Oliver and Oberthaler, Markus},
  journal = {Rev. Mod. Phys.},
  volume = {78},
  issue = {1},
  pages = {179--215},
  numpages = {0},
  year = {2006},
  month = {Feb},
  publisher = {American Physical Society},
  doi = {10.1103/RevModPhys.78.179},
  url = {https://link.aps.org/doi/10.1103/RevModPhys.78.179}
}

@article{BM2011,
 author = {R. Bistritzer and A. MacDonald},
 journal = {Proc. Nat. Acad. Sci},
 pages = {12233–
12237},
 title = {Moir\'e bands in twisted double-layer graphene},
 url = {https://doi.org/10.1080/23746149.2018.1473052},
 volume = {108},
 year = {2011}
}

@misc{lape2025realizationcharacterizationallbandsflatelectrical,
      title={Realization and characterization of an all-bands-flat electrical lattice}, 
      author={Noah Lape and Simon Diubenkov and L. Q. English and P. G. Kevrekidis and Alexei Andreanov and Yeongjun Kim and Sergej Flach},
      year={2025},
      eprint={2508.13571},
      archivePrefix={arXiv},
      primaryClass={cond-mat.mes-hall},
      url={https://arxiv.org/abs/2508.13571}, 
}

@article{SPKW-inpreparation,
  title = {Flat band lattices and nonlinear bound states in discrete nonlinear wave equations},
  author = {Shi, C. and Parker, R. and Kevrekidis, P. G. and Weinstein, M.I. },
  journal = {in preparation},
}

@article{clugru:hal-03268886,
  TITLE = {{General Construction and Topological Classification of All Magnetic and Non-Magnetic Flat Bands}},
  AUTHOR = {C{\u a}lug{\u a}ru, Dumitru and Chew, Aaron and Elcoro, Luis and Xu, Yuanfeng and Regnault, Nicolas and Song, Zhi-Da and Bernevig, B. Andrei},
  URL = {https://hal.science/hal-03268886},
  JOURNAL = {{Nature Phys.}},
  VOLUME = {18},
  NUMBER = {2},
  PAGES = {185-189},
  YEAR = {2022},
  DOI = {10.1038/s41567-021-01445-3},
  HAL_ID = {hal-03268886},
  HAL_VERSION = {v1},
}

@article{leykam2018artificial,
 author = {Leykam, Daniel and Andreanov, Alexei and Flach, Sergej},
 doi = {10.1080/23746149.2018.1473052},
 journal = {Adv. Phys.: X},
 number = {1},
 pages = {1473052},
 publisher = {Taylor \& Francis},
 title = {Artificial flat band systems: from lattice models to experiments},
 url = {https://doi.org/10.1080/23746149.2018.1473052},
 volume = {3},
 year = {2018},
}

@article{drouot_lyman:2025,
 author = {Drouot, A. and Lyman, C.},
 journal = {https://arxiv.org/abs/2410.02092},
 title = {Band spectrum singularities for Schrödinger operators},
 year = {2025},
}

@article{leykam2013flat,
 author = {Leykam, Daniel and Flach, Sergej and Bahat-Treidel, Omri and Desyatnikov, Anton S.},
 doi = {10.1103/PhysRevB.88.224203},
 issue = {22},
 journal = {Phys. Rev. B},
 month = dec,
 numpages = {6},
 pages = {224203},
 publisher = {American Physical Society},
 title = {Flat band states: Disorder and nonlinearity},
 url = {https://link.aps.org/doi/10.1103/PhysRevB.88.224203},
 volume = {88},
 year = {2013},
}

@article{danieli2024flat,
  url = {https://doi.org/10.1515/nanoph-2024-0135},
  title = {Flat band fine-tuning and its photonic applications},
  author = {Carlo Danieli and Alexei Andreanov and Daniel Leykam and Sergej Flach},
  pages = {3925--3944},
  volume = {13},
  number = {21},
  journal = {Nanophotonics},
  doi = {doi:10.1515/nanoph-2024-0135},
  year = {2024},
  eprint = {2403.17578},
  archivePrefix = {arXiv},
  primaryClass = {physics.optics},
}

@article{rhim2021singular,
 author = {Rhim, Jun-Won and Yang, Bohm-Jung},
 title = {Singular flat bands},
 journal = {Advances in Physics: X},
 volume = {6},
 number = {1},
 pages = {1901606},
 year = {2021},
 publisher = {Taylor \& Francis},
 doi = {10.1080/23746149.2021.1901606},
 url = {https://doi.org/10.1080/23746149.2021.1901606},
}

@article{vicencio2021photonic,
  author = {Rodrigo A. Vicencio Poblete},
  title = {Photonic flat band dynamics},
  journal = {Advances in Physics: X},
  volume = {6},
  number = {1},
  pages = {1878057},
  year = {2021},
  publisher = {Taylor & Francis},
  doi = {10.1080/23746149.2021.1878057},
  url = {https://doi.org/10.1080/23746149.2021.1878057},
  eprint = {https://doi.org/10.1080/23746149.2021.1878057}
}

@article{weimann2016transport,
 author = {Weimann, Steffen and Morales-Inostroza, Luis and Real, Basti\'{a}n and Cantillano, Camilo and Szameit, Alexander and Vicencio, Rodrigo A.},
 doi = {10.1364/OL.41.002414},
 journal = {Opt. Lett.},
 month = jun,
 number = {11},
 pages = {2414--2417},
 publisher = {OSA},
 title = {Transport in Sawtooth photonic lattices},
 url = {http://ol.osa.org/abstract.cfm?URI=ol-41-11-2414},
 volume = {41},
 year = {2016},
}

@article{vicencio2013discrete,
  title = {Discrete flat-band solitons in the kagome lattice},
  author = {Vicencio, Rodrigo A. and Johansson, Magnus},
  journal = {Phys. Rev. A},
  volume = {87},
  issue = {6},
  pages = {061803},
  numpages = {5},
  year = {2013},
  month = {Jun},
  publisher = {American Physical Society},
  doi = {10.1103/PhysRevA.87.061803},
  url = {https://link.aps.org/doi/10.1103/PhysRevA.87.061803}
}

@article{Law,
  title = {Localized structures in kagome lattices},
  author = {Law, K. J. H. and Saxena, Avadh and Kevrekidis, P. G. and Bishop, A. R.},
  journal = {Phys. Rev. A},
  volume = {79},
  issue = {5},
  pages = {053818},
  numpages = {13},
  year = {2009},
  month = {May},
  publisher = {American Physical Society},
  doi = {10.1103/PhysRevA.79.053818},
  url = {https://link.aps.org/doi/10.1103/PhysRevA.79.053818}
}

@book{malomed2013spontaneous,
  editor       = {Malomed, Boris A.},
  title        = {Spontaneous Symmetry Breaking, Self-Trapping, and Josephson Oscillations},
  series       = {Progress in Optical Science and Photonics},
  volume       = {1},
  publisher    = {Springer Berlin Heidelberg},
  address      = {Berlin, Heidelberg},
  year         = {2013},
  edition      = {1st},
  isbn         = {978-3-642-21206-2},
  doi          = {10.1007/978-3-642-21207-9},
  pages        = {xvii + 707},
  note         = {339 black\-and\-white illustrations; 1 in colour},
}

@article{guzman2014experimental,
 author = {Guzm{\'a}n-Silva, D and Mej{\'\i}a-Cort{\'e}s, C and Bandres, M A and Rechtsman, M C and Weimann, S and Nolte, S and Segev, M and Szameit, A and Vicencio, R A},
 journal = {New J. Phys.},
 number = {6},
 pages = {063061},
 title = {Experimental observation of bulk and edge transport in photonic {Lieb} lattices},
 url = {http://stacks.iop.org/1367-2630/16/i=6/a=063061},
 volume = {16},
 year = {2014},
}

@article{mukherjee2015observation,
 author = {Mukherjee, Sebabrata and Thomson, Robert R.},
 doi = {10.1364/OL.40.005443},
 journal = {Opt. Lett.},
 month = dec,
 number = {23},
 pages = {5443--5446},
 publisher = {OSA},
 title = {Observation of localized flat-band modes in a quasi-one-dimensional photonic rhombic lattice},
 url = {http://ol.osa.org/abstract.cfm?URI=ol-40-23-5443},
 volume = {40},
 year = {2015},
}

@article{vicencio2015observation,
 author = {Vicencio, Rodrigo A. and Cantillano, Camilo and Morales-Inostroza, Luis and Real, Basti\'an and Mej\'{\i}a-Cort\'es, Cristian and Weimann, Steffen and Szameit, Alexander and Molina, Mario I.},
 doi = {10.1103/PhysRevLett.114.245503},
 issue = {24},
 journal = {Phys. Rev. Lett.},
 month = jun,
 numpages = {5},
 pages = {245503},
 publisher = {American Physical Society},
 title = {Observation of Localized States in {Lieb} Photonic Lattices},
 url = {http://link.aps.org/doi/10.1103/PhysRevLett.114.245503},
 volume = {114},
 year = {2015},
}

@article{kajiwara2016observation,
 author = {Kajiwara, Sho and Urade, Yoshiro and Nakata, Yosuke and Nakanishi, Toshihiro and Kitano, Masao},
 doi = {10.1103/PhysRevB.93.075126},
 issue = {7},
 journal = {Phys. Rev. B},
 month = feb,
 numpages = {7},
 pages = {075126},
 publisher = {American Physical Society},
 title = {Observation of a nonradiative flat band for spoof surface plasmons in a metallic {Lieb} lattice},
 url = {https://link.aps.org/doi/10.1103/PhysRevB.93.075126},
 volume = {93},
 year = {2016},
}

@article{nguyen2018symmetry,
 author = {Nguyen, H.S. and Dubois, F. and Deschamps, T. and Cueff, S. and Pardon, A. and Leclercq, J.-L. and Seassal, C. and Letartre, X. and Viktorovitch, P.},
 doi = {10.1103/PhysRevLett.120.066102},
 issue = {6},
 journal = {Phys. Rev. Lett.},
 month = feb,
 numpages = {6},
 pages = {066102},
 publisher = {American Physical Society},
 title = {Symmetry Breaking in Photonic Crystals: On-Demand Dispersion from Flatband to Dirac Cones},
 url = {https://link.aps.org/doi/10.1103/PhysRevLett.120.066102},
 volume = {120},
 year = {2018},
}

@article{ma2020direct,
  title = {Direct Observation of Flatband Loop States Arising from Nontrivial Real-Space Topology},
  author = {Ma, Jina and Rhim, Jun-Won and Tang, Liqin and Xia, Shiqi and Wang, Haiping and Zheng, Xiuyan and Xia, Shiqiang and Song, Daohong and Hu, Yi and Li, Yigang and Yang, Bohm-Jung and Leykam, Daniel and Chen, Zhigang},
  journal = {Phys. Rev. Lett.},
  volume = {124},
  issue = {18},
  pages = {183901},
  numpages = {7},
  year = {2020},
  month = {May},
  publisher = {American Physical Society},
  doi = {10.1103/PhysRevLett.124.183901},
  url = {https://link.aps.org/doi/10.1103/PhysRevLett.124.183901}
}

@article{masumoto2012exciton,
 author = {Masumoto, Naoyuki and Kim, Na Young and Byrnes, Tim and Kusudo, Kenichiro and L{\"o}ffler, Andreas and H{\"o}fling, Sven and Forchel, Alfred and Yamamoto, Yoshihisa},
 journal = {New J. Phys.},
 number = {6},
 pages = {065002},
 title = {Exciton--polariton condensates with flat bands in a two-dimensional kagome lattice},
 url = {http://stacks.iop.org/1367-2630/14/i=6/a=065002},
 volume = {14},
 year = {2012},
}

@article{whittaker2018exciton,
  title = {Exciton Polaritons in a Two-Dimensional Lieb Lattice with Spin-Orbit Coupling},
  author = {Whittaker, C. E. and Cancellieri, E. and Walker, P. M. and Gulevich, D. R. and Schomerus, H. and Vaitiekus, D. and Royall, B. and Whittaker, D. M. and Clarke, E. and Iorsh, I. V. and Shelykh, I. A. and Skolnick, M. S. and Krizhanovskii, D. N.},
  journal = {Phys. Rev. Lett.},
  volume = {120},
  issue = {9},
  pages = {097401},
  numpages = {6},
  year = {2018},
  month = {Mar},
  publisher = {American Physical Society},
  doi = {10.1103/PhysRevLett.120.097401},
  url = {https://link.aps.org/doi/10.1103/PhysRevLett.120.097401}
}

@article{wang2022observation,
  title={Observation of inverse Anderson transitions in Aharonov-Bohm topolectrical circuits},
  author={Wang, Haiteng and Zhang, Weixuan and Sun, Houjun and Zhang, Xiangdong},
  journal={Physical Review B},
  volume={106},
  number={10},
  pages={104203},
  year={2022},
  publisher={APS}
}

@article{zhou2023observation,
  title={Observation of flat-band localization and topological edge states induced by effective strong interactions in electrical circuit networks},
  author={Zhou, Xiaoqi and Zhang, Weixuan and Sun, Houjun and Zhang, Xiangdong},
  journal={Physical Review B},
  volume={107},
  number={3},
  pages={035152},
  year={2023},
  publisher={APS}
}

@article{chasemayoral2024compact,
  title = {Compact localized states in electric circuit flat-band lattices},
  author = {Chase-Mayoral, Carys and English, L. Q. and Lape, Noah and Kim, Yeongjun and Lee, Sanghoon and Andreanov, Alexei and Flach, Sergej and Kevrekidis, P. G.},
  journal = {Phys. Rev. B},
  volume = {109},
  issue = {7},
  pages = {075430},
  numpages = {9},
  year = {2024},
  month = {Feb},
  publisher = {American Physical Society},
  doi = {10.1103/PhysRevB.109.075430},
  url = {https://link.aps.org/doi/10.1103/PhysRevB.109.075430},
  eprint = {2307.15319},
  archivePrefix = {arXiv},
  primaryClass = {cond-mat.mes-hall}
}

@article{xia2016demonstration,
 author = {Xia, Shiqiang and Hu, Yi and Song, Daohong and Zong, Yuanyuan and Tang, Liqin and Chen, Zhigang},
 doi = {10.1364/OL.41.001435},
 journal = {Opt. Lett.},
 month = apr,
 number = {7},
 pages = {1435--1438},
 publisher = {OSA},
 title = {Demonstration of flat-band image transmission in optically induced Lieb photonic lattices},
 url = {http://ol.osa.org/abstract.cfm?URI=ol-41-7-1435},
 volume = {41},
 year = {2016},
}

@book{reed1978,
  author    = {Michael Reed and Barry Simon},
  title     = {Methods of Modern Mathematical Physics, Volume IV: Analysis of Operators},
  publisher = {Academic Press},
  year      = {1978},
  address   = {New York}
}

@article{Weinstein1986,
  author    = {Michael I. Weinstein},
  title     = {Lyapunov Stability of Ground States of Nonlinear Dispersive Evolution Equations},
  journal   = {Communications on Pure and Applied Mathematics},
  volume    = {39},
  number    = {1},
  pages     = {51--67},
  year      = {1986},
  doi       = {10.1002/cpa.3160390103}
}

@article{Weinstein1999,
  author    = {Michael I. Weinstein},
  title     = {Excitation Thresholds for Nonlinear Localized Modes on Lattices},
  journal   = {Nonlinearity},
  volume    = {12},
  number    = {3},
  pages     = {673--691},
  year      = {1999},
  doi       = {10.1088/0951-7715/12/3/314},
  url       = {https://doi.org/10.1088/0951-7715/12/3/314}
}

@book{Weinstein2015,
  author    = {Michael I. Weinstein},
  title     = {Localized states and their dynamics in the nonlinear Schroedinger / Gross-Pitaeveskii equation: Analysis and Applications},
  journal   = {Frontiers in Applied Dynamics: Reviews and Tutorials},
  volume    = {3},
  year      = {2015},
 publisher = {Springer}
}

@book{Golub-VanLoan2013,
  author    = {G.H. Golub and C.F. Van Loan},
  title     = {Matrix Computations, 4th ed.},
  year      = {2015},
 publisher = {Johns Hopkins, 2013}
}

@article{Kuchment2005,
  author  = {Peter Kuchment},
  title   = {Quantum Graphs II: Some Spectral Properties of Quantum and Combinatorial Graphs},
  journal = {Journal of Physics A: Mathematical and Theoretical},
  volume  = {38},
  number  = {22},
  pages   = {4887--4900},
  year    = {2005},
  doi     = {10.1088/0305-4470/38/22/013},
  url     = {https://doi.org/10.1088/0305-4470/38/22/013}
}

@article{Leykam2018,
  author  = {Daniel Leykam and Alexander Andreanov and Sergej Flach},
  title   = {Artificial Flat Band Systems: From Lattice Models to Experiments},
  journal = {Advances in Physics: X},
  volume  = {3},
  number  = {1},
  pages   = {1473052},
  year    = {2018},
  doi     = {10.1080/23746149.2018.1473052},
  url     = {https://doi.org/10.1080/23746149.2018.1473052}
}

@article{Grillakis1987,
  author  = {Manoussos Grillakis and Jalal Shatah and Walter Strauss},
  title   = {Stability Theory of Solitary Waves in the Presence of Symmetry, I},
  journal = {Journal of Functional Analysis},
  volume  = {74},
  number  = {1},
  pages   = {160--197},
  year    = {1987},
  doi     = {10.1016/0022-1236(87)90044-9},
  url     = {https://doi.org/10.1016/0022-1236(87)90044-9}
}

@article{Grillakis1988,
  author  = {Manoussos G. Grillakis},
  title   = {Linearized Instability for Nonlinear Schrödinger and Klein-Gordon Equations},
  journal = {Communications on Pure and Applied Mathematics},
  volume  = {41},
  number  = {6},
  pages   = {747--774},
  year    = {1988},
  doi     = {10.1002/cpa.3160410602},
  url     = {https://doi.org/10.1002/cpa.3160410602}
}

@article{Jones1988,
  author  = {C.K.R.T. Jones},
  title   = {Instability of standing waves for nonlinear Schr\"odinger-type equations},
  journal = {Ergod. Th. \& Dynam. sys.},
  volume  = {8},
  pages   = {119-138},
  year    = {1988}
}

@article{fefferman2012honeycomb,
  title = {Honeycomb lattice potentials and {Dirac} points},
  author = {C. L. Fefferman and M. I. Weinstein},
  journal = {Journal of the American Mathematical Society},
  volume = {25},
  number = {4},
  pages = {1169-1220},
  year = {2012}
}

@article{keller2018spectral,
  title = {Spectral band degeneracies of $\frac{\pi}{2}$-rotationally invariant periodic {Schr\"{o}dinger} operators},
  author = {R. T. Keller and J. L. Marzuola and B. Osting and M. I. Weinstein},
  journal = {Multiscale Modeling \& Simulation},
  volume = {16},
  number = {4},
  pages = {1684-1731},
  year = {2018},
  publisher = {SIAM}
}

@article{Mielke1991,
  author    = {Mielke, Andreas},
  title     = {Ferromagnetic ground states for the Hubbard model on line graphs},
  journal   = {Journal of Physics A: Mathematical and General},
  year      = {1991},
  volume    = {24},
  number    = {2},
  pages     = {L73--L77},
  doi       = {10.1088/0305-4470/24/2/005}
}

@article{Mielke1991a,
  author    = {Mielke, Andreas},
  title     = {Ferromagnetism in the Hubbard model on line graphs and further considerations},
  journal   = {Journal of Physics A: Mathematical and General},
  year      = {1991},
  volume    = {24},
  number    = {14},
  pages     = {3311--3321},
  doi       = {10.1088/0305-4470/24/14/018}
}

@article{Kollar2020,
  author    = {Koll{\'a}r, Alicia J. and Fitzpatrick, Mattias and Sarnak, Peter and Houck, Andrew A.},
  title     = {Line‐Graph Lattices: Euclidean and Non‐Euclidean Flat Bands, and Implementations in Circuit Quantum Electrodynamics},
  journal   = {Communications in Mathematical Physics},
  volume    = {376},
  number    = {3},
  pages     = {1909--1956},
  year      = {2020},
  doi       = {10.1007/s00220-019-03645-8}
}

@article{kim2025real,
  title     = {Real space decay of flat band projectors from compact localized states},
  author    = {Kim, Yeongjun and Flach, Sergej and Andreanov, Alexei},
  journal   = {arXiv preprint arXiv:2510.17258v1},
  year      = {2025}
}

\clearpage

\appendix 

\begin{center}
    {\bf End Matter}
\end{center}

\noindent {\bf Framework for stability / instability analysis.}\  
 A nonlinear standing wave $\psi^\nu e^{-iEt}$ is a critical point in $l^2(\mathbb L)$ of $\mathcal{H}[u]$ subject to the constraint $\mathcal{N}[u]=\nu$, i.e. for some $E_{\rm nl}=E_{\rm nl}(\nu)$, $\psi^\nu$ satisfies $\delta\mathcal{E}[\psi^\nu]=0$, where $\mathcal{E}[u]\equiv \mathcal{H}[u]-E_{\rm nl}(\nu)\mathcal{N}[u]$. A strategy to explore whether a state $\psi^\nu$ is stable is to examine whether $\psi^\nu$ is a local constrained maximizer \cite{Weinstein1986, Weinstein1999, Grillakis1987}.
  To this end, we let $\psi\in l^2(\mathbb{L})$ such that $\mathcal{N}[\psi] = \nu$, and choose $\theta_\star$ so that $\|\psi - e^{i\theta_\star}\psi^\nu\| =  {\rm dist}\big(\psi, \mathcal{O}_{\psi^\nu}\big)$. Writing $u+iv =\psi -e^{i\theta_\star}\psi^\nu$ we have 
$\mathcal{H}[\psi]-\mathcal{H}[\psi^\nu] \approx \left\langle L_+u,u\right\rangle + \left\langle L_-v,v\right\rangle$,
where, for the case of MCS states under power law nonlinearities, $L_\pm$ are displayed in \eqref{eq:Lpm}.
The optimality of $\theta_*(t)$ implies $\left\langle \psi^\nu,v\right\rangle= 0$, and $\mathcal{N}[\psi] = \mathcal{N}[\psi^\nu]$ implies $\left\langle \psi^\nu,u\right\rangle = 0$. As operators on $l^2(\mathbb L)$, $L_-$ is negative semi-definite and $L_+$ is indefinite and has at least one positive eigenvalue. 
Hence, the state $\psi^\nu$ is a constrained local maximizer and is (orbitally Lyapunov) stable if the above quadratic form in $(u,v)$, subject to the orthogonality constraint on $u$, is negative definite. Since the MCS $\psi^\nu$ itself is an eigenvector of $L_+$ with a strictly positive eigenvalue, it follows from the orthogonality constraint that a sufficient condition for $\psi^\nu$ being a constrained local maximizer is that all other eigenvalues of $L_+$ are strictly negative, i.e., $L_+$ has only one non-negative eigenvalue.


On the other hand, conditions on the spectra of  $L_+$ and $L_-$, which imply the existence of unstable eigenvalues of the linearized time-evolution, were established in \cite{grillakis1988,jones1988} in the context of solitary standing waves of the continuum {\it focusing} NLS equation.  Adapting the results of \cite{grillakis1988} to MCS states of defocusing $\mathbb L-$DNLS, we have that: a sufficient condition for linearized instability (a time exponentially growing linearized solution) is that the number of strictly positive eigenvalues of  $L_+$ is strictly larger than one. 

\noindent {\bf Use of the properties of MCS states in the spectral analysis.}
The proof of Theorem \ref{theorem: power law nonlinearity stability} reduces the spectral analysis of $L_+$ to  the family of rank one operators $\{L_{+,\omega^n}\}_{n=s/2}$; see \eqref{eq:L+rank1}. $L_{+,\omega^n}$ is given in terms of $e_n\in  l_{\omega^n}^2(\mathbb L)$ ($n\ne s/2$), which is localized on the support of $\psi_a$. 
The states $e_n$ can be characterized using the following observation: since $e_n\in l_{\omega^n}^2(\mathbb L)$ and ${\rm supp}(e_n) = {\rm Orb}_\mathfrak{S}(x)$ for some $x\in\mathbb L$ with $|{\rm Orb_\mathfrak S}(x)| = s$, $e_n$ is generated by $e_n(x)$, because $e_n(\mathfrak{S}x) = \omega^{-n}e_n(x)$. This suggests that each $e_n$ acts as a ``wave function" and forms a Fourier-type basis on ${\rm supp}(\psi_a)$. 

Calculating the resolvent $L_{+,\omega^n}$  is tractable via the Sherman-Morrison-Woodbury identity \cite{Golub-VanLoan2013}, which yields an expression for $(L_{+,\omega^n}-\lambda {\rm I})^{-1}$  a rank one perturbation of the resolvent of the bulk operator, which is unbounded only on the continuous spectrum.
For each $n\ne s/2$, there is a single pole of the resolvent (point eigenvalue of $L_{+,\omega^n}$) at the zero, $\lambda_n(|a|^{2\sigma})$, of the function $F_n(\lambda, |a|^{2\sigma})$, which is displayed in Eq.~(\ref{equation: Sherman-Morrison --- invertibility}).
%
%

Fix $n\ne s/2$.
We study the eigenvalue curves 
$|a|^{2\sigma}\in(0,\infty)\mapsto \lambda_n\big(|a|^{2\sigma}\big)$. Note first that  $F_n\big(|a|^{2\sigma},\lambda=0\big)\to-2\sigma<0$ as $|a|^{2\sigma}\to\infty$. Next, by the functional calculus of self-adjoint operators \cite{reed1978}, we have
\[
\lim_{|a|^{2\sigma}\to 0}F_n(|a|^{2\sigma},\lambda = 0) = 1- (2\sigma+1) \big\|P_{\{E_F\}}(-\Delta_{\mathbb L})e_n\big\|^2.
\]
Hence, if $\big\|P_{\{E_F\}}(-\Delta_{\mathbb L})e_n\big\|^2<(2\sigma+1)^{-1}$, then  $F_n(0^+,0)>0$.  Therefore, by continuity, there exists $a_0(n)>0$, for which $\lambda_n(|a_0(n)|^{2\sigma})=0$. Implicit differentiation shows that $|a|^{2\sigma}\mapsto \lambda_n(|a|^{2\sigma})$ is strictly increasing once $\lambda_n(|a|^{2\sigma})\geq 0$, which implies the uniqueness of the zero crossing point $a_0(n)$. A detailed and rigorous analysis is carried out in \cite{SPKW-inpreparation}.

Therefore, as $a$ increases through $a_0(n)$, the eigenvalue $\lambda_n(|a|^{2\sigma})$ transitions from the interval $-|a|^{2\sigma}<\lambda<0$ to positive values. It follows that if condition (\ref{equation: stability criterion}) is satisfied then
 for $0<a<a_\star := \min_{n\neq s/2}a_0(n)$, the $s-1$ eigenvalues
 $\lambda_{n}(|a|^{2\sigma})$, with $n\ne s/2$,  are strictly negative and we have stability of the MCS state $\psi_a$.

\noindent {\bf Global study of bifurcations.
}\  For the diamond lattice, $\mathbb D$, the MCS state is a dipole supported on two sites. It can be proved that $\sigma_{\mathbb D}^{\rm cr} = \infty$.  As seen in Fig.~\ref{fig:BD}(a), for $\sigma=1$,
the dipole state is stable for $\mathcal N<\nu_\star\approx 2.61$ 
[branch \#1], and becomes unstable for $\mathcal{N}>\nu_\star$, which is consistent with Theorem \ref{theorem: power law nonlinearity stability}.  At $\nu_\star$, $L_+$, the linearized operator about the dipole, has a nontrivial nullspace, which triggers a bifurcation. The corresponding zero eigenstate exhibits even horizontal reflection symmetry, a symmetry not shared by the anti-symmetric dipole. The resulting bifurcating branch therefore breaks the symmetry type of the MCS state, and emerges as an asymmetric  dipole [branch \#2]. The states on this new branch are no longer compactly supported.
Fig.~\ref{fig:BD}(a) displays additional solutions (``single-site'' [branch \#3] and ``double'' [branch \#5]) which are not compactly supported and bifurcate from the
 quadratic band; both are unstable.  

For the Kagom{\'e} lattice, $\mathbb K$, the MCS state interacts with other solution branches via bifurcation when $(\sigma = 1<1.268\approx\sigma^{\rm cr}_{\mathbb K})$; see Fig.~\ref{fig:BD}(b). The MCS hexagonal states [branch \#1] are stable at small power. At $\nu_\star$, this stability is lost through a symmetry-breaking bifurcation associated with a nontrivial kernel of $L_+$. The resulting asymmetric solution [branch \#2] eventually connects to the single-site state [branch \#3], which is unstable until this bifurcation point.
This behavior was numerically reported in~\cite{vicencio2013discrete} and is mathematically explained herein.

A somewhat analogous scenario occurs for the checkerboard lattice, $\mathbb Ch$, in Fig.~\ref{fig:BD}(d). When $\sigma = 0.6<\sigma_{\mathbb Ch}^{\rm cr}\approx 0.829$, the MCS square state [branch \#1] loses stability at a power $\nu_\star$. Similar to the Kagom{\'e} lattice, the corresponding zero crossing of an eigenvalue of $L_+$ induces bifurcation of an intermediate branch \#2, which then connects to the single-site state branch \#3. 
In contrast to $\mathbb K$, the bifurcation at $E_{\rm nl}\approx 2.3$ is subcritical; both branches \#1 for $\nu > \nu_\star$ and $\#2$ for $\nu < \nu_\star$ are unstable. 



Consistent with Theorem 1, we observe instability of all hexagonal MCS states for quartic nonlinearity $(\sigma = 1.5>\sigma^{\rm cr}_{\mathbb K}\approx 1.268)$ in $\mathbb K$ (Fig.~\ref{fig:BD}(c)), as well as of all square MCS states for cubic nonlinearity ($\sigma = 1>\sigma_{\mathbb Ch}^{\rm cr}\approx0.829$) in $\mathbb Ch$ (Fig.~\ref{fig:BD}(e)).


\bigskip

\begin{figure*}[h]
    \centering
    \begin{minipage}{\linewidth}
        \centering
        \begin{overpic}[width=0.95\linewidth]{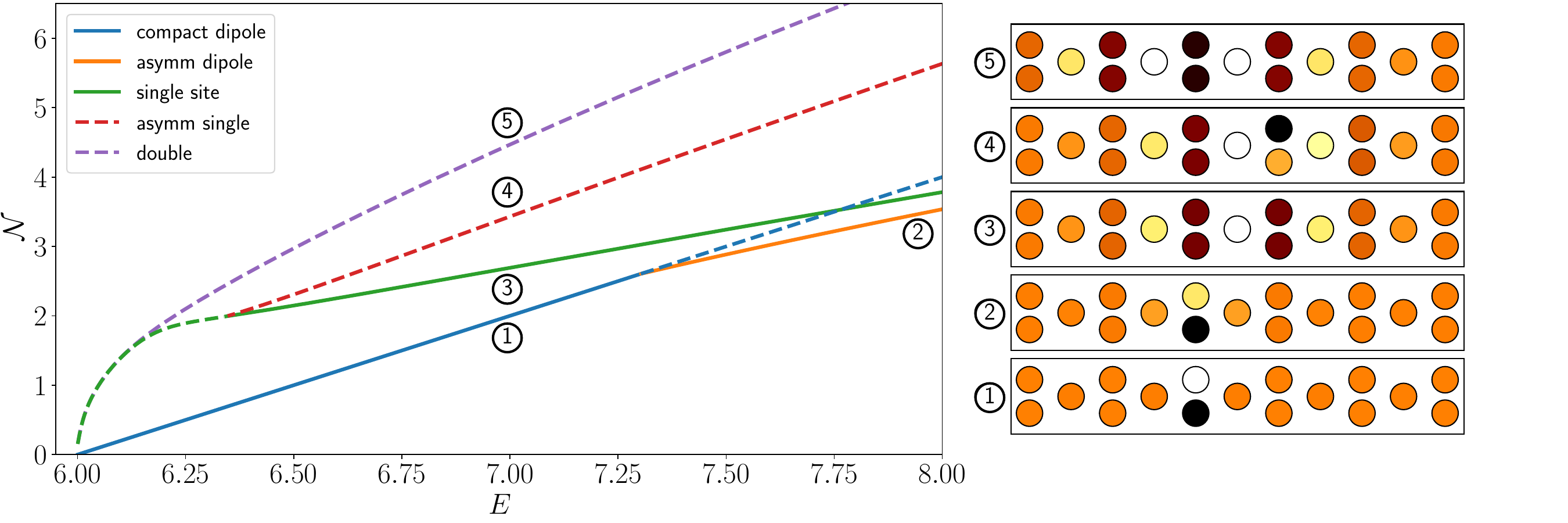}
            \put(-2,32){\textbf{(a)}}
        \end{overpic}
    \end{minipage}

    \vspace{3mm}

    \begin{minipage}{\linewidth}
        \centering
        \begin{overpic}[width=0.95\linewidth]{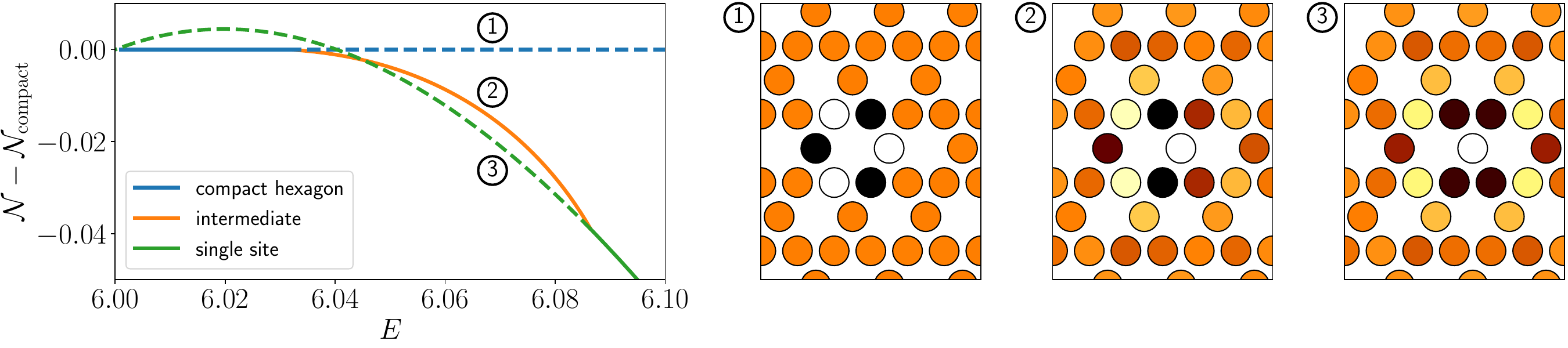}
            \put(-2,21){\textbf{(b)}}
        \end{overpic}
    \end{minipage}

    \vspace{3mm}

    \begin{minipage}{\linewidth}
        \centering
        \begin{overpic}[width=0.95\linewidth]{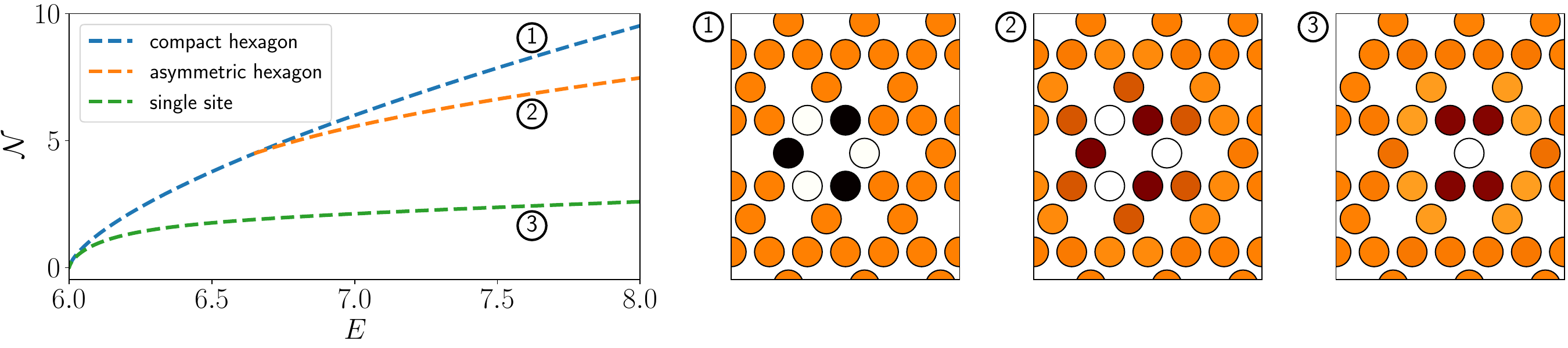}
            \put(-2,21){\textbf{(c)}}
        \end{overpic}
    \end{minipage}

    \vspace{3mm}

    \begin{minipage}{\linewidth}
        \centering
        \begin{overpic}[width=0.95\linewidth]{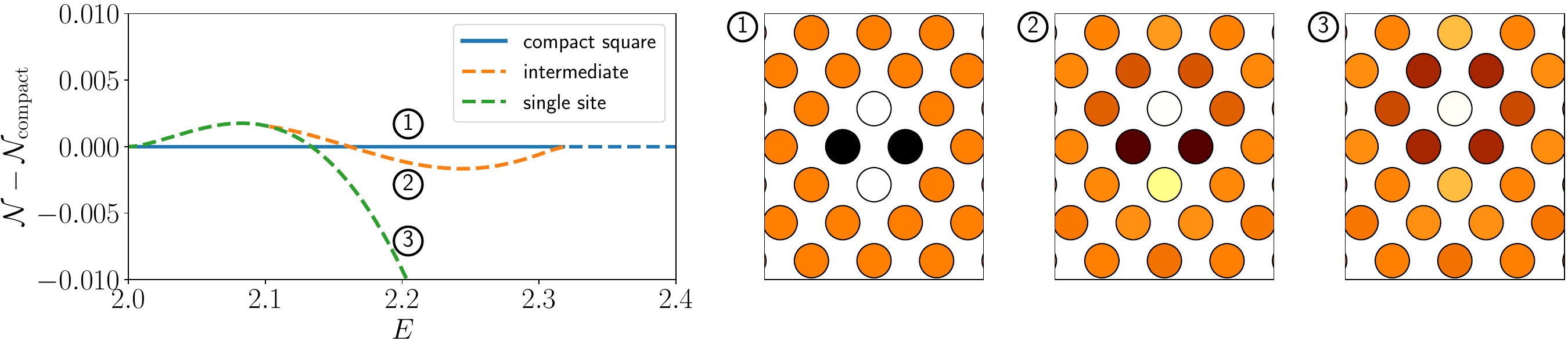}
            \put(-2,21){\textbf{(d)}}
        \end{overpic}
    \end{minipage}

    \vspace{3mm}

    \begin{minipage}{\linewidth}
        \centering
        \begin{overpic}[width=0.95\linewidth]{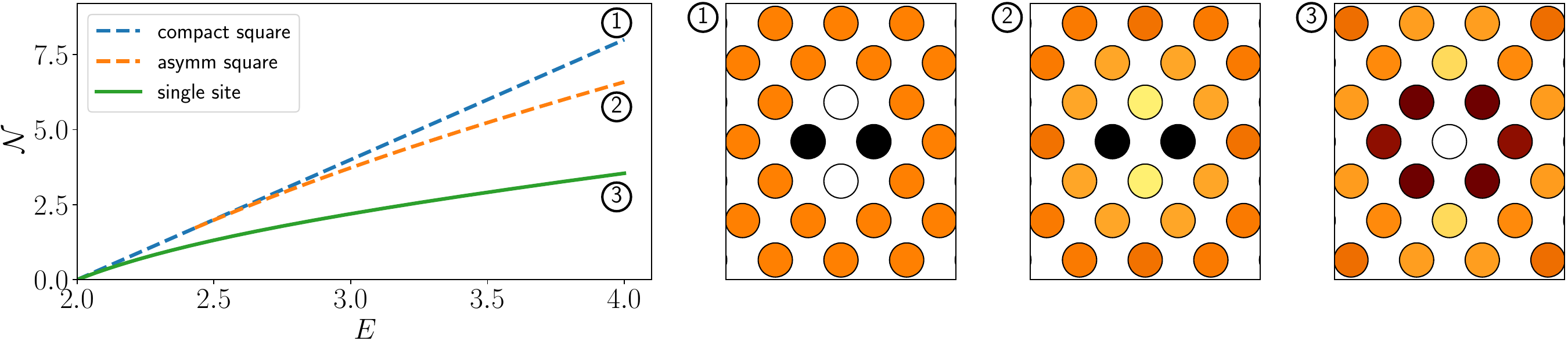}
            \put(-2,21){\textbf{(e)}}
        \end{overpic}
    \end{minipage}
    \caption{Bifurcation diagrams plotting power $\mathcal{N}$ vs. nonlinear frequency $E$ for diamond lattice, Kagom{\'e} lattice, and checkerboard lattice. (a) Diamond lattice $\mathbb D-$DNLS with $\sigma =1$. (b) and (c), Kagom{\'e} lattice $\mathbb K-$DNLS with $\sigma = 1 < \sigma_{\mathbb K}^{\rm cr}$, and  $\sigma = 1.5 > \sigma_{\mathbb K}^{\rm cr}$, respectively. (d) and (e), Checkerboard lattice $\mathbb Ch-$DNLS with  $\sigma = 0.6 < \sigma^{\rm cr}_{\mathbb Ch}$, and  $\sigma = 1 > \sigma^{\rm cr}_{\mathbb Ch}$, respectively. Solid lines indicate spectral stability, dotted lines indicate spectral instability. (The vertical axis for (b) and (d) is the difference between the power of the solution and the power of the compact solution, which was chosen for ease of visualization). }
    \label{fig:BD}
\end{figure*}

\bigskip

\end{document}